\documentclass[epj]{svjour}
\usepackage{epsfig}
\newcommand{\be}{\begin{equation}}
\newcommand{\ee}{\end{equation}}
\newcommand{\bea}{\begin{eqnarray}}
\newcommand{\eea}{\end{eqnarray}}
\newcommand{\ba}{\begin{array}}
\newcommand{\ea}{\end{array}}

\begin{document}
\title{Active Brownian Motion Models and Applications to Ratchets}

\author{Alessandro Fiasconaro\inst{1}\inst{2}\inst{4} \thanks{email: afiasconaro@gip.dft.unipa.it} \and Werner Ebeling\inst{1}\inst{3} \and Ewa Gudowska--Nowak \inst{1}\inst{4}}
\institute{Mark Kac Complex Systems Research Center, Jagellonian
University, Reymonta 4, 30-059 Krak\'ow, Poland. \and Dipartimento
di Fisica e Tecnologie Relative\footnote{Group of
Interdisciplinary Physics, http://gip.dft.unipa.it} and CNISM,
Universit\`a di Palermo, Viale delle Scienze, I-90128 Palermo,
Italy\and Institute of Physics, Humboldt University Berlin,
Newtonstr. 15, 12489 Berlin, Germany \and Marian~Smoluchowski
Institute of Physics, Jagellonian University, Reymonta 4, 30-059
Krak\'ow, Poland.}
\date{Received: date / Revised version:}
%
\abstract{We give an overview over recent studies on the model of
Active Brownian Motion (ABM) coupled to reservoirs providing free
energy which may be converted into kinetic energy of motion.
First, we present an introduction to a general concept of active
Brownian particles which are capable to take up energy from the
source and transform
part of it in order to perform various activities. In the second
part of our presentation we consider applications of ABM to
ratchet systems with different forms of differentiable potentials.
Both analytical and numerical evaluations are discussed for three
cases of sinusoidal, staircase-like and Mateos ratchet potentials,
also with the additional loads modeled by tilted potential
structure. In addition, stochastic character of the kinetics is
investigated by considering perturbation by Gaussian white noise
which is shown to be responsible for driving the directionality of
the asymptotic flux in the ratchet. This \textit{stochastically
driven directionality} effect is visualized as a strong
nonmonotonic dependence of the statistics of the right versus left
trajectories
 of motion leading to a net current of particles.
  Possible applications of the
ratchet systems to molecular motors are also briefly discussed.
} 
\maketitle
\section{Introduction}
\label{intro} The study of mechanical systems with support
of energy goes back to the investigations of Helmholtz and
Rayleigh on the origin of sustained oscillations.
Generalizations of the "active friction" introduced by
Rayleigh found many applications including the concept of
the Active Brownian Motion (ABM) which extends the notion of
standard Brownian motion as studied by Einstein,
Smoluchwski, Fokker, Planck and others \cite{Smol16} to the
field of driven motions
\cite{Kl94,EbSo05,SchwEbTi98,EbSchwTi99,ErEbSchiSchw00}
including a developing theory
of swarming motions \cite{Vi01,MiCa02,Schw03,EbAPP7,EbScEJP7}.

In this work we first introduce a model of Langevin
dynamics coupled to energy depot dynamics and study basic
properties. The inclusion of the depot dynamics should model
the general observation that external energy which is needed for acceleration of motion is in most cases
stored first in a depot or reservoir (a kind of an energy tank) and, only in a second step, becomes converted by a kind of motor into motion. This storage and subsequent conversion of the energy into mechanical work is modeled here in the simplest possible way by a balance equation.
Next we study more specific applications to
transport problems on Hamiltonian ratchets and
discuss applications to molecular
motors.

Since the fundamental work of Marian Smoluchowski \cite{Smol16}
the problem of transport on ratchets is under a constant debate.
Some of the most interesting applications are related to
biological problems
(see e.g. J\"{u}licher and Prost,\cite{Jul95,Jul97}).
Here we will discuss several problems related to the ABM
of particles on ratchet potentials. Further we will discuss
 possible use of ABM in modelling the functioning of ATP and ADP in
cells and the related transport mechanisms, having in mind
possible applications to biological systems as e.g. proton pumps
and electron pumps.
Some further relations with the stepping motor described e.g. by Bier
\cite{Bier03,Bier07} are also  proposed.

\section{Model of Brownian motion coupled to energy reservoirs}

\subsection{Coupling between Langevin dynamics and energy depot dynamics}

We postulate  dynamics of Brownian particles as determined by the
Langevin equation according to a model proposed by Schweitzer et
al. \cite{SchwEbTi98}:
\begin{equation}
  \label{langev-or}
  \dot{\bf r}_i={\bf v}_i\,; \qquad
  m\dot{\bf v}_i + \nabla U({\bf r}) = {\bf F}_i  + \sqrt{2D}{\bf \xi}(t)
  \label{evolution}
\end{equation}
where $U(r)$  is the potential of the conservative forces and
$\sqrt{2D}{\bf \xi(t)}$ is a stochastic force with strength $D$ and a
$\delta$-correlated time dependence:
\begin{equation}
  \label{stoch}
  \langle {\xi}_i (t) \rangle=0 \,; \qquad
  \langle {\xi}_i(t){\xi}_j (t')\rangle=
  \delta(t-t') \delta_{ij}
\end{equation}
The driving  forces on the RHS of Eq.(\ref{evolution}) are expressed in the form \cite{SchwEbTi98}
\begin{equation}
  {\bf F}_i =  -m \gamma{\bf v} + m d e(t) {\bf v}
  \label{friction}
\end{equation}
where the first term stands for the dissipative force and the second is responsible for the acceleration of movement due to the conversion of  the depot energy into kinetic energy of motion. Under the equilibrium conditions, the friction coefficient $\gamma$ (here defined as the velocity dependent function $\gamma({\bf v})$) is a constant  directly related to the noise strength $D$ by the Einstein relation $D=mk_BT\gamma$.
Depot energy dissipation and coupling of energy reservoir to the kinetic degrees of freedom give
rise to the time-dependent balance equation for $e(t)$
 \begin{equation}
   \dot{e}(t) = q - c e(t) - d e(t) {\bf v^2}
 \label{ereservoir}
 \end{equation}
 Here $q$ is the take-up of energy term and $ce(t)$ describes the internal dissipation in the reservoir which is assumed to be proportional to the depot energy $e(t)$. The conversion of depot into kinetic energy of motion is controlled by the rate $d$ and depends (quadratically) on the actual velocity of the particle. The overall time variation of the mechanical energy of the particle can be derived from Eqs.(1-4) and yields
  \begin{eqnarray}
 \frac{d}{dt}\left (\frac{m}{2}v^2+ U({\bf r})\right )={\bf F_i}{\bf v_i}+\sqrt{2D}\xi(t) {\bf v_i}=\nonumber \\
  =-\frac{d}{dt}me(t) + m (q-ce(t))-m{\bf v_i^2}\gamma+\sqrt{2D}\xi(t) {\bf v_i}
 \end{eqnarray}

In an adiabatic approximation (assumed rapid relaxation of
variations in $e(t)$ to its stationary value, i.e. $\dot{e}(t)=0$)
we may substitute the energy $e(t)$ in Eq. (\ref{friction}) by its
stationary value and get a general form for the dissipative force
 \be {\bf F_i} = - m\gamma(v_i^2)
{\bf v_i} \ee The function $\gamma(v^2)$ denotes a
velocity-dependent friction, which in our model has a negative
part. Accordingly, the depot model \cite{SchwEbTi98,ErEbSchiSchw00} for the energy supply leads to

\begin{equation}
  \gamma({\bf v}^2)= \left(\gamma - \frac{d q}{c + d v^2}\right)
\end{equation}
where $c,d,q$ are certain positive constants characterizing the
energy flows from the depot to the particles. Depending on the
parameters $\gamma$, $c$, $d$ and $q$ the dissipative force
function may have one zero at ${\bf v} = 0$ or two more zeros at
\begin{equation}
  {\bf v}_0^2=\beta,
\label{v-0}
  \end{equation}
when the bifurcation parameter
\begin{equation}
   \beta = \frac{q}{\gamma} -\frac{c}{d}
\end{equation}
becomes positive. In this case
 a finite characteristic velocity $v_0$ exists which
determines an attractor of motion and one speaks of an active
Brownian-particles motion. For $|{\bf v}| < v_{0}$, the
dissipative force is positive, i.e.\ the particle is provided with
additional free energy. Hence, slow particles are accelerated,
while the motion of fast particles becomes damped. Note that in
the case of thermal equilibrium systems we have $\gamma({\bf
v^2})=\gamma_{0}={\rm const}$. The adiabatic treatment of the
equations proposed by Schweitzer et al. \cite{SchwEbTi98} found
many applications to problems of active Brownian motions as e.g.
swarm dynamics \cite{SchwEbTi01,ErEbMi05,Tilch,SchwEbTi00}.
However, in some cases, and in particular for applications to
ratchet problems the adiabatic approximation understood as mere adiabatic elimination of fast variables
from the dynamic equations of motion may lead to improper
conclusions as will be discussed in the forthcoming paragraphs.

\subsection{Free Brownian particles and the action of a constant external force}

Let us first discuss the free motion of active particles in a one or
two-dimensional space, $\vec r=\{x_{1},x_{2}\}$. Under the condition of a quasi-stationary depot, $\dot{e}(t)=0$, the long time (stationary) probability distribution function of
the Fokker-Planck equation

\begin{eqnarray}
\frac{\partial P(\vec r,\vec v, t)}{\partial t}+ \vec v \frac{\partial P(\vec r,\vec v, t)}{\partial \vec r} +
\frac{\nabla U({\bf r})}{m}\frac{\partial P(\vec r,\vec v, t)}{\partial \vec v}\nonumber \\
=\frac{\partial}{\partial \vec v}\left [ \gamma(\vec r, \vec v)\vec v P(\vec r,\vec v, t)+D_v \frac{\partial P(\vec r,\vec v, t)}{\partial \vec v}\right ]
\end{eqnarray}
corresponding to the Langevin equation Eq.(1) with $U(\vec r)\equiv 0$ can be easily found and reads
\cite{ErEbSchiSchw00}
\begin{equation}
  P_0(\vec v) = C \left(1 + \frac{d  v^2}{c}\right)^{(q/2D_v)}
  \exp\left[ -\frac{\gamma} {2D_v} \, v^2\right],
  \label{P_0}
\end{equation}
where $D_v = D/m^2$ (cf. Eqs. (1,2)) stands for the diffusion coefficient in the velocity space. Accordingly,  the mean square displacement
\begin{eqnarray}
<(\vec r_1(t) - \vec r_1(0))^2 > =\nonumber \\
\int_0^{\infty}dt_1\int_0^{\infty}dt_2<\vec v(t_1)\vec v(t_2)>
\end{eqnarray}
can be evaluated and in the limit of a strong, supercritical influx of energy ($q\rightarrow\infty$)  and for times much longer than $v_0^2/2D_v$ leads to an approximate expression
\begin{eqnarray}
  <(\vec r_1(t) - \vec r_1(0))^2 > = \frac{v_0^4}{D_v} (2 t)= 2D_{eff}t =\nonumber \\
  =\frac{2v_0^4m^2}{D}t.
  \label{mean}
\end{eqnarray}
So far the usual Brownian motion and the active Brownian
motion seem to behave in a quite similar way: we see that the dispersion of the displacement grows linearly with time resembling typical character of the diffusive motion  with $D_{eff}=v_0^4/D_v$ . However, the prefactor $2D_{eff}$ is completely different from the value $2k_BT(m\gamma)^{-1}$ which rules the relation $ <(\vec r_1(t) - \vec r_1(0))^2 >=2k_BT(m\gamma)^{-1}$ for  a standard Brownian diffusion.
 In case of ABM, due to the inverse-proportional dependence on  the noise intensity $D$ (cf. Eq. (1)),  weak noise gives rise to large mean square displacement which is one of the peculiar properties of the motion \cite{EbSo05}.\\

From now on we will consider only one-dimensional problems. First we
analyze the deterministic equations for several instructive
special cases (for simplicity of derivations we assume $m=1$):
 \begin{equation}
   \ddot{x}(t) + U'(x) +[\gamma- de(t)] \dot{x}(t) = F_0
 \label{a_eq1}
 \end{equation}
 \begin{equation}
   \dot{e}(t) - q + ce(t) + de(t) \dot{x}^2(t) =0
 \label{a_eq2}
 \end{equation}
Here $F_0$ stands for a possible tilt of the potential. As a
zeroth-order approximation we first neglect in Eq.(\ref{a_eq1}) the
term  $U'(x)=0$ and we adiabatically eliminate the energy term
from the second equation by approximating $\dot e =0$:
\begin{eqnarray}
\dot{x}=v\nonumber \\
\dot{v}=-[\gamma- de(t)]v + F_0\nonumber \\
\dot{e}(t)=0\rightarrow e=\frac{q}{c+d v^2}
\end{eqnarray}
 When
treating the set of the above equations by use of the
equilibrium condition (vanishing of the dissipation term in the
second equation) which requires $e_0= \gamma/d$ we assume that
the stability flux of energy counterbalances the friction term:
\begin{eqnarray}
\dot{x}=v\nonumber \\
\dot{v}= F_0\nonumber \\
e=e_0=\frac{\gamma}{d}=\frac{q}{c+dv_0^2}
\label{equil}
\end{eqnarray}
The above set of equation implies now that $F_0=0.$

 We
will take later this exact solution (with $F_0=0$ and $U'(x)=0$) as the starting point of a
perturbation theory. The bifurcation parameter of our problem is
   $\beta$.
For $\beta \geq 0$, the system is driven to non-equilibrium states (note that this case requires pumping of energy from the reservoir)
and has all together three stationary states of the velocity: $v=0$,
$v=v_0$ and $v=-v_0$.

\begin{figure}[htbp]
 \begin{center}
  \includegraphics[angle=-90,width=8.5cm]{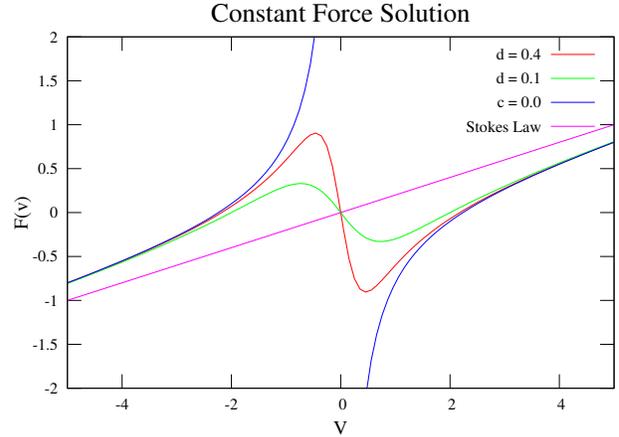}
\caption{Stationary solution: The force $F(v)=F_0=-a$ as a
function of the velocity for $\gamma = 0.2$ and 2 fixed
$d$-values ($d=0.4$ and $d=0.1$ with $c=0.1$) and the
asymptotic curve with $c=0$. The comparison is made with
the Stokes law $F(v)=\gamma v$. The solution to Eq. (\ref{forcev}) representing free
motion corresponds to the roots on the axis $F(v)=0$ }
  \label{fcost}
 \end{center}%
\end{figure}

We consider now the case of a constant tilt by means of the
additive force
 \be
    F_0 = - a.
 \ee
Here $a$ is the slope of an equivalent potential $U_0 (x) = a x$.
We will mostly focus on a positive slope $a
> 0, F_0 <0$. Let us first consider the case in the absence of
an additional ratchet potential i.e. $U(x) =0$. This problem still
admits an exact solution. Without an energy flux $q$ from the
reservoir, the particle would fall down (if $a>0$ from right to
left). Including the reservoir provides the possibility of uphill
motions. The condition of stationary motion under the action of
this force leads to the cubic equation
 \be
   \gamma d v_0^3 -F_0d v_0^2 +(c \gamma - q d) v_0 -F_0 c=0
   \label{forcev}
 \ee
The solutions may be found graphically (see Fig. \ref{fcost}). If
we assume that dissipation $c$ and the slope $a$ are so small that
$a c=-F_0 c$ may be neglected, the solution reads
 \be
  v_0= \frac{F_0}{2\gamma} \pm \sqrt
{\frac{F_0^2}{4\gamma^2}+\frac{q}{\gamma}-\frac{c}{d}}
 \ee
Altogether, in the general case we have a cubic equation for the
stationary velocities, which is easy to solve numerically, and in
some special cases, also analytically. Following the structure of
the linear potential $U_0$, the downhill motion exists in all
cases and for our standard choice $a> 0; F_0<0$ the downhill
motion is directed to the left. Remarkably,  in the case of positive
energy input $q >0$ also a stationary uphill motion may exist
$v_0>0$, provided the force driving downhill is not too large. For
example, if $a = 1; F_0=-1$, the trivial downhill solution is $v_0
=-5$ and the stable uphill solution is $v_0 =+0.5$ for $d=1$; for
$d=0.3$ no uphill solution exists. It is interesting to mention that
the uphill motion may exist even without any ratchet effects provided
the driving is sufficiently strong. Still the question remains, what is the influence of
the ratchet potential on the directionality of transport and what is the mechanical efficiency of
the system.\\
This brief examination  allows us to conclude about the minimum set-up
conditions for construction of a ratchet-type potential in which the
unidirectional current can be obtained:\\
(i) the average value of the flatter slope of the potential should be in the range
where the uphill motion
is possible, and \\
(ii) the average value of the steeper slope should not allow the uphill motion.\\
Under these conditions the particle can go uphill from left to
right and, since the motion backwards is not possible, we get a
unidirectional movement. We will use this construction as a rule
of thumb in order to find the conditions for directed transport in one dimensional periodic structures.

\subsection{Effect of white noise in the mechanical
equations with constant force}

\begin{figure}[htbp]
 \begin{center}
  \includegraphics[angle=-90,width=8.5cm]{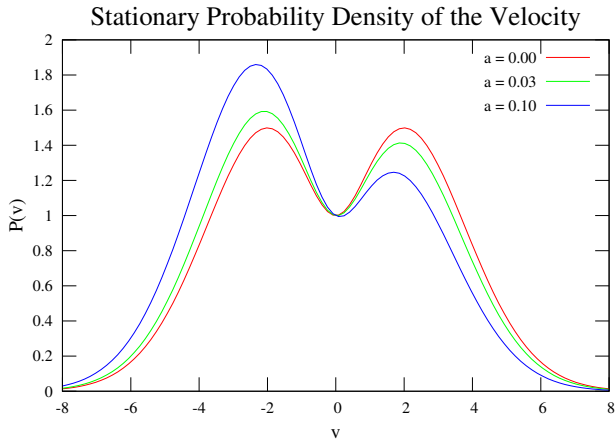}
\caption{Example of the asymmetric probability distribution
of the asymptotic velocity under a ratchet in the presence
of a tilt for different values of the constant force
intensity $a$. The tilt of the potential gives rise to the
asymmetry of the distribution.}
  \label{asymdistr}
 \end{center}
\end{figure}

Having in mind biological applications where typically some
transfer from chemical to mechanic or electric energy appears in
the presence of random fluctuations, we will study now ratchets
which are connected to an energy reservoir under the influence of
noise. As a generalization of Eq.(\ref{a_eq1})

 \be
   \frac{dv(t)}{dt}+ \gamma v(t) + U'(x) = F_0 + d e(t) v(t) + \sqrt{2D}\xi(t)
   \nonumber
 \ee
The equation for the energy depot remains instead exactly
Eq.(\ref{a_eq2}). For special case when  $U'(x) = 0$, $F_0 = 0$ and $\dot{e}=0$,
the corresponding Fokker-Planck equation may be solved and the
solution is given by Eq. (\ref{P_0}). When including a constant tilt
$F_0 = - a < 0$, the stationary Fokker-Planck equation may also be solved
exactly. The solution gives now an asymmetrical distribution with
two maxima corresponding to the deterministic stable flux
velocities (see Fig. \ref{asymdistr}):
 \be P_0(v) = C
\exp\left[-\frac{ \gamma v^2}{2 D} -a v + \frac{q}{2D}
\log(1+\frac{d}{c} v^2) \right].
 \ee
The shape of the stationary probability density demonstrated in Fig.2 shows that under the action of a constant force directed to the left ($a>0$), flows in both directions are possible, however with a different probability: Flows directed to the right are less probable than those oriented to the left.
\section{Ratchets coupled to energy sources}

\subsection{Models of inertia ratchets}

\begin{figure}
 \begin{center}
  \includegraphics[angle=-90, width=8.5cm]{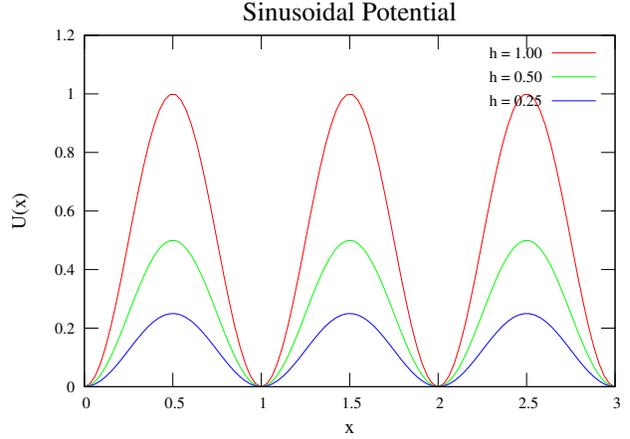}
  \caption{Sinusoidal potential with parameter $h=1.0, 0.5, 0.25$.}
  \label{pot_sin}
 \end{center}%
\end{figure}

In a series of recent papers \cite{Tilch,SchwEbTi00} various aspects of the ABM energetics in an external potential have been analyzed. In order to investigate further the motion of an ensemble of pumped Brownian particles in periodic fields, we consider two simple models of smooth 1-dim potentials:
 the symmetric sinusoidal potential
 (see Fig.\ref{pot_sin}):
\begin{equation}
   U(x) = \frac{h}{2} [1 - \cos(2\pi x)]
 \label{pot}
\end{equation}
and the asymmetric ratchet potential (See Fig.\ref{pot_sin} and Fig.\ref{pot_mat}) introduced by Mateos and Machura \cite{Mateos01,Machura}:
\bea
  U(x) = h \{0.499 - 0.453\{\sin[2 \pi (x + 0.1903)] + \nonumber \\
   + 0.25 \sin[4 \pi (x + 0.1903)]\}
   \}.
 \label{pot}
 \eea
 We first address the problem of the classical deterministic dynamics of a particle in those potentials. We construct the bifurcation diagram and identify the origin of the current. By analyzing stochastic influences, we detect noise-induced directionality of the current.

\subsection{Sinusoidal Ratchets in the trapped regime}

We study first oscillations around $x=0$. In this case the
potential can be approximated by a parabolic one
 \be
   U(x) \simeq h \pi^2 x^2 = \frac{1}{2}  \omega_0^2 x^2.
 \ee
and for a positive value of the bifurcation parameter $\beta > 0$
the system displays self-oscillating solutions. By  assuming their
form as
 \begin{equation}
   v(t)= a sin(\omega t - \alpha),
 \end{equation}
and substituting into the set of evolution equations, we may
determine the amplitude $a$. For small $\beta$ a standard
derivation of a periodic solution yields $a=(2\beta)^{1/2}$ and
$\omega=\omega_0$.

\begin{figure}[htbp]
 \begin{center}
  \includegraphics[angle=-90,width=8.5cm]{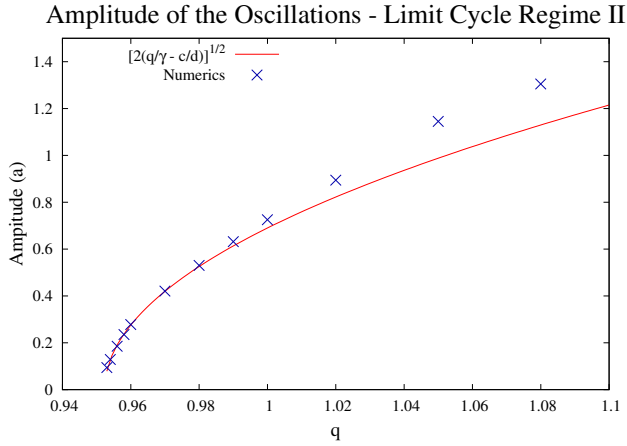}
 \caption{Amplitude of the limit cycle as a function of driving
parameter $q$ for different values beyond the critical values
(symbols) in comparison with the theoretical result (full line).We
used a ratchet with $h=1$, driving parameters $\gamma=0.2$,
$c=0.1$, $d=0.021$. For amplitudes larger than around half of the
maximal amplitude, the theory breaks down due to significant
deviations of the sine-potential from a parabola.}
  \label{amp}
 \end{center}%
\end{figure}

In Fig. \ref{amp} the plot of the amplitude of the asymptotic
velocity $a$ as a function of the parameter $q$ is displayed. We
see that small amplitudes follow predicted root law whereas
divergence between analytical and numerical results at higher
values of $q$ are due  to a breakdown of the parabolic
approximation of the potential.
For higher values we observe a bifurcation from a limit cycle  to
an open trajectory. The transition to open trajectories is
expected in the region where the kinetic energy exceeds the
maximal potential energy:
 \begin{equation}
 \frac{1}{2} v_0^2 = \frac{1}{2} \left( \frac{q}{\gamma}-\frac{c}{d} \right) \geq
   h \end{equation}
Consequently, in what follows, we use the energy transfer
parameter $d$ as an order parameter in our study. The equality
sign ($=$) in the above expression defines the value of the energy
exchange parameter $d$ giving rise to a bifurcation between stable
oscillations and the flux regime. We denote this value $d_E$.
 \be
   d_E= \frac{d_c}{1- d_c h/c}
   \label{de}
 \ee
where $d_c=\gamma c / q$. The above expression, as based on the parabolic
approximation of the potential, gives only a rough
estimation of $d_E$ and is no longer valid in the
dynamical region where the transition from stable oscillations
(trapped trajectories) to the flux regime takes place. As a
consequence, the numerically evaluated parameter ($d_{E,n}$)
differs significantly from the value predicted in this
approximation ($d_{E,a}$) (See Fig.\ref{bifurcation}).

For particles  entrapped in the potential well and
performing sustained oscillations one can derive an
analytical expression for the value of energy in the
energy-reservoir. The starting point for the derivation is
an assumption that at a moderate energy transfer
parameterized by $d$, a confined closed orbit occurs in a
potential well. Let's assume that this trajectory is
characterized by a sinusoidal velocity:
 \begin{equation}
   \dot{x}_{\infty}(t) = a sin(\omega t),
 \label{a_eq3}
 \end{equation}
After substituting $ \dot{x}_{\infty}(t)$ in Eq.(\ref{a_eq2}) and integration, the analytical expression for the depot energy $e(t)$ takes on the form:
 \begin{equation}
   e_{\infty}(t) = (qW(t) + e_0 ) e^{-\eta(t)}
   \label{en}
 \end{equation}
 where
 \begin{equation}
   W(t)=\int_0^t e^{\eta(t')} dt'
 \label{a_eq5}
 \end{equation}
and
 \begin{equation}
   \eta(t)=(c+da^2/2)t-\frac{da^2}{4\omega}sin(2\omega t).
 \end{equation}
 The integral Eq.(\ref{a_eq5}) can be evaluated in the long time limit and the result implemented again to the formula
 Eq.(\ref{en}) leading to an asymptotic expression for the depot energy
 \begin{equation}
   e_{\infty}(t)=\frac{q}{c+da^2/2} \exp{\left( \frac{da^2}{4\omega} sin(2\omega t) \right)}.
  \label{a_eq7}
 \end{equation}
which averaged over the time becomes:
 \begin{equation}
   e_{\infty} = \langle e_{\infty}(t) \rangle_{\tau} =\frac{q}{c+da^2/2} \left\langle \exp{\left(\frac{da^2}{4\omega} sin(2\omega t)\right)}\right\rangle_{\tau}.
 \end{equation}
Here brackets represent the time average
\begin{equation}
\langle f(t) \rangle_{\tau}\equiv\lim_{\tau\rightarrow \infty}\frac{1}{\tau}\int^{\tau}_0f(t)dt
\end{equation}

\begin{figure}[htbp]
 \begin{center}
  \includegraphics[angle=-90,width=8.5cm]{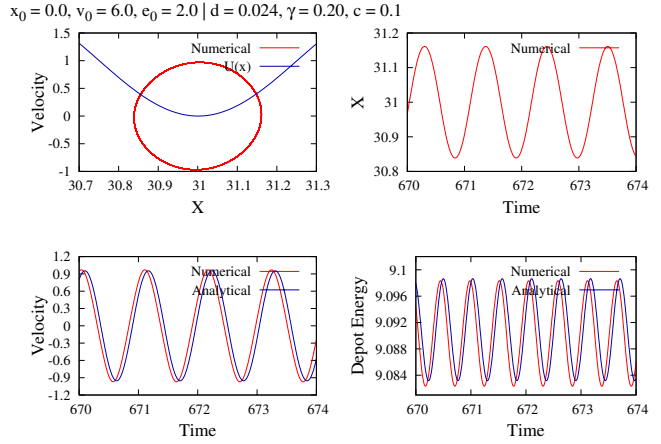}
 \caption{Asymptotic behavior of the system in the trapped
case with limit cycle ($d_c<d<d_E$). $a=
\sqrt{\frac{2q}{\gamma}-\frac{2c}{d}}$. Graphs of
time-dependent energy represent variation of the depot
energy $e(t)$.}
  \label{cycle}
 \end{center}%
\end{figure}

\begin{figure}[htbp]
 \begin{center}
 \includegraphics[angle=-90,width=8.5cm]{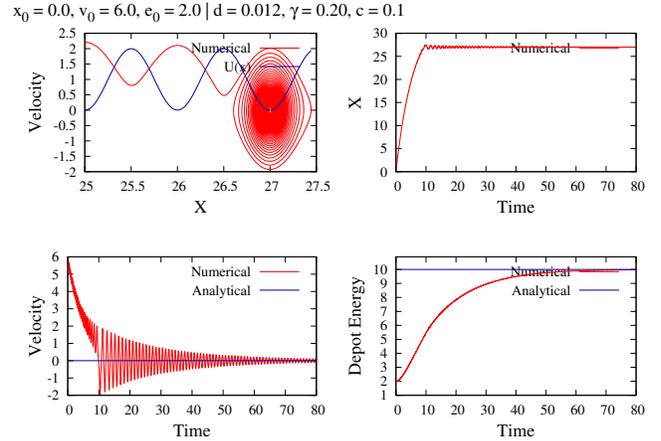}
 \caption{Asymptotic behavior of the system in the trapped
case with zero limit ($d \leq d_c$)and imaginary parameter
$a$. Graphs of time-dependent energy represent variation of
the depot energy $e(t)$.}
  \label{zero}
 \end{center}%
\end{figure}

In a first approximation\footnote{This approximation holds
only for $\frac{da^2}{4\omega} \ll 1$, which is a valid
value for our case. A more general value can be found as
follow: we have that $\langle e^{\frac{da^2}{4\omega}
sin(2\omega t)}\rangle \epsilon [e^{-\frac{da^2}{4\omega}},
e^{\frac{da^2}{4\omega}}]$, so we can put as approximation
the mean value between the two extremes of the exponential
function: $\Rightarrow \langle e^{\frac{da^2}{4\omega}
sin(2\omega t)}\rangle \approx cosh
(\frac{da^2}{4\omega})$. In this more general case, the
relation between the parameters is: $\frac{q}{c + d
a^2/2}cosh (\frac{da^2}{4\omega}) \approx
\frac{\gamma}{d}$. This relation is not simple to invert
and also requires $\omega$ as an input value to obtain
$a$.}
 we can use
 \begin{equation}
   \left\langle \exp{\left(\frac{da^2}{4\omega} sin(2\omega t)\right)}\right\rangle_{\tau} \approx 1.
 \end{equation}
due to  the average $\langle sin(2\omega t)\rangle_{\tau}=0$. This means
 that
 \begin{equation}
   e_{\infty} \approx \frac{q}{c+da^2/2} .
 \label{a_eq10}
 \end{equation}
which differs from Eq.(\ref{equil}) by a factor $1/2$. On
the other hand, by assuming close-to-equilibrium condition
(dissipation term vanishing over the time average) in Eq.
(\ref{a_eq1}) we can evaluate:
 \begin{equation}
   \langle \gamma - d e_{\infty}(t) \rangle_{\tau} = 0
 \end{equation}
This means
that:
 \begin{equation}
   e_{\infty} = \frac{\gamma}{d}.
 \label{a_eq12}
 \end{equation}
Comparing Eq.(\ref{a_eq1}) and Eq.(\ref{a_eq12}), we have a
relation between the parameters of the equations and the
amplitude $a$ of the limit velocity of the particle:
 \begin{equation}
   \frac{q}{c+da^2/2}=\frac{\gamma}{d} \Rightarrow a= \sqrt{\frac{2q}{\gamma}-\frac{2c}{d}}=\sqrt{2 \beta}.
   \label{a_eq13}
 \end{equation}
This predicted value for $a$ is very close to that found by
the numerical solution of the equations. From the former
equation is it possible to evaluate the limit threshold
giving the stable oscillation of the system. In fact, by
putting $a=0$ we obtain the value
 \begin{equation}
   d_c=\frac{\gamma c}{q}.
   \label{a_eq14}
 \end{equation}
To resume the present result, we have, for $d_c < d < d_E$:
 \begin{itemize}
  \item
    stable oscillation of $v(t)$ with amplitude $a$
  \item
    stable oscillation of $x(t)$ in a certain well of the potential with amplitude $a/2\pi\omega$
  \item
    oscillating stationary behavior for $e_{\infty}(t)$ with frequency $\omega_e =
    2\omega$,
 \end{itemize}
while for $d \leq d_c$ we have:
 \begin{itemize}
  \item
    limit value $x(t)$ in a certain final position $x_f$
  \item
    decreasing of $v(t)$ ($a=0$)
  \item
 damped oscillating behavior with saturation for
$e_{\infty}(t)$ (see Eq.(\ref{a_eq7}) for $a^2 = 0$)
 \end{itemize}
Results of numerical evaluation, together with the
analytical result Eq.(\ref{a_eq7}), are plot in two
figures below (Figs.\ref{cycle}, \ref{zero}).
  As it can be
deduced from the simulations, motion in the periodic
ratchet potential leads to oscillatory behavior of the
energy $e(t)$. That refrains us from using strictly the
elimination scheme based on the assumption $\dot{e}(t)=0$.
Instead, the periodic variation of $e(t)$ leads to an
average (over time) constant value of $e_{\infty}$ that
differs from the value predicted in Eq.(\ref{equil}).

\subsection{Sinusoidal ratchets in the flux regime}
We will try first a perturbation theory developing the
expression of the velocity of our system:
 \begin{eqnarray}
   v=v_0 + v_1(t) + ...;\nonumber \\
   e= e_0 + e_1(t) + ...
 \end{eqnarray}
For large driving and small forces ($U'(x) \approx 0$) the
particles move as free and there are two attractors of the
velocity
 \begin{eqnarray}
   v_0^+= \sqrt{\beta},\nonumber \\
    v_0^- =- \sqrt{\beta}\nonumber \\
     e_0 = \frac{\gamma}{d}
 \end{eqnarray}
We take these solutions as the first term in a perturbation
series. Inserting this $zero$- approximation into the Eq.
\ref{a_eq1}, we have:
 \be
   \dot{v}_1(t) = \frac{d v_1(x)}{dx}\frac{d x}{dt} \approx \frac{d v_1(x)}{dx} v_0 = -U'(x)
 \ee
>From which, integrating in $x$, we have the solution:
 \be
   v_1(x) v_0 \approx  -U(x) + Const
 \ee
and therefore, up to the first order:
 \be
   v_{(1)}(x)=v_0+v_1(x) \approx  v_0 -\frac{U(x)}{v_0} + \frac{Const}{v_0}.
 \ee
The choice of the constant, is made in order to have a the
mean value of the potential equal to $0$ (See Eq.
(\ref{pot})). The value of the constant is then $Const =
0.499 h$. With this choice we obtain, up to the 1st order of
approximation:
 \be
   v_{(1)}(x) \approx  v_0 -\frac{U(x)}{v_0} +
   \frac{0.499 h}{v_0}.
 \label{approx}
 \ee
The above expression is valid, in principle, for any shape
of differentiable potential ratchets.

\begin{figure}[htbp]
 \begin{center}
  \includegraphics[angle=-90,width=8.5cm]{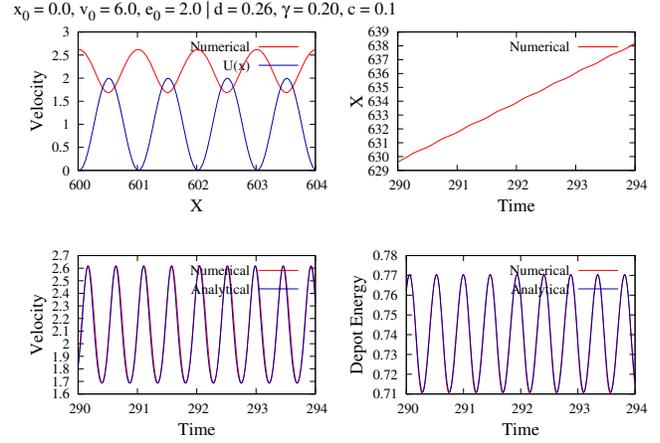}
\caption{Case of asymptotical flux. Numerical and
analytical behaviors are quite superimposed. Differently
from the trapped case, the energy oscillates mainly with
the same frequency of the velocity.}
  \label{flux}
 \end{center}%
\end{figure}

\begin{figure}[htbp]
 \begin{center}
\includegraphics[angle=-90,width=8cm]{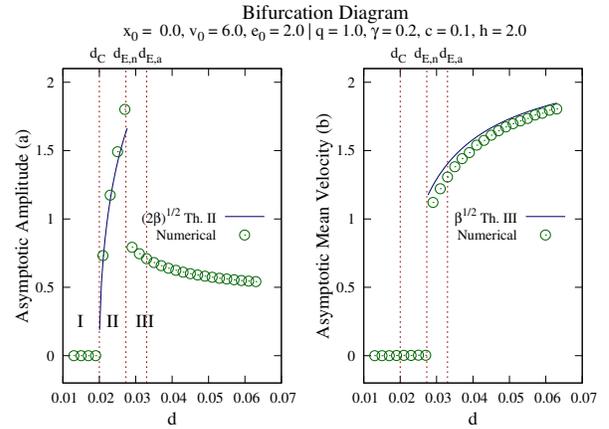}
\caption{Bifurcation diagram for the amplitude (left hand
side) and the mean value (right) of the asymptotic
velocity. The comparison between the predicted values and
the measured ones are plotted. The three region I, II, III,
represents the different dynamical regimes as a function of
the order parameter $d$. The indicated threshold values
$d_{E,n}$ $d_{E,a}$ represents respectively the numerical
and analytical evaluation (assuming the harmonic potential
profile) of the escape bifurcation parameter $d_E$ (see
Eq.\ref{de})}
  \label{bifurcation}
 \end{center}%
\end{figure}

Taking into account the Eq. \ref{approx}, we can make the
assumption for our asymptotic velocity:
 \begin{equation}
   \dot{x}_{\infty}(t) = b + a sin(\omega t),
 \label{eq3}
 \end{equation}
the analytical expression for the $e(t)$ is then:
 \begin{equation}
   e_{\infty}(t) = (qZ(t) + e_0 e^{-2dba/\omega} ) e^{-\xi'(t)}
 \end{equation}
 where
 \begin{equation}
   Z(t)=\int_0^t e^{\xi(t')} dt'
 \label{eq5}
 \end{equation}
and
 \be
   \xi(t)=(c + d b^2+da^2/2)t- \frac{\zeta(t)}{\omega}.
 \ee
with
 \be
   \zeta(t)=2 d b a cos(\omega t)+\frac{d a^2}{4} sin(2 \omega t).
 \ee
Because the term $(c + d b^2+da^2/2)$ is greater than zero,
the equation has a non vanishing and non diverging
asymptotic behavior. The asymptotic expression we can
extract from the equation is now:
 \begin{equation}
   e_{\infty}(t)=\frac{q}{c + d b^2+da^2/2} e^{\zeta(t)/\omega}.
 \end{equation}
The expression for $e_{\infty}(t)$ presents in the
exponential two oscillating functions with two different
frequencies ($\omega$ and $2\omega$). Because the
coefficient of the cosine function is greater than that of
the sine, the observable frequency in the
reservoir energy is the same than the velocity one, while
in the trapped case we have only the frequency $2\omega$
for the asymptotic energy. Fig.\ref{flux} shows the
behavior of this expression compared with that one obtained
by numerical evaluation. The agreement is very good. The
parameter $a$ and $b$ can be evaluated by means of the
comparison with the Eqs.(\ref{eq3}) and (\ref{approx}), while
$\omega$ has been extracted from the numerical evaluation
of the limit velocity $v_{\infty}(t)$.

\subsection{Influence of external forces and noise}
Assuming that the external force is different from zero, we
put a certain bias to the right or left direction. A particle which
is able to go uphill at the cost of the supply from the reservoir energy
may convert this energy into mechanical or electrical energy and perform work. This simple motor device is of particular interest in nanobiotechnology, where the reservoir
energy is in most realistic cases just the chemical energy of reaction.

\begin{figure}[htbp]
 \begin{center}
  \includegraphics[angle=-90, width=8.5cm]{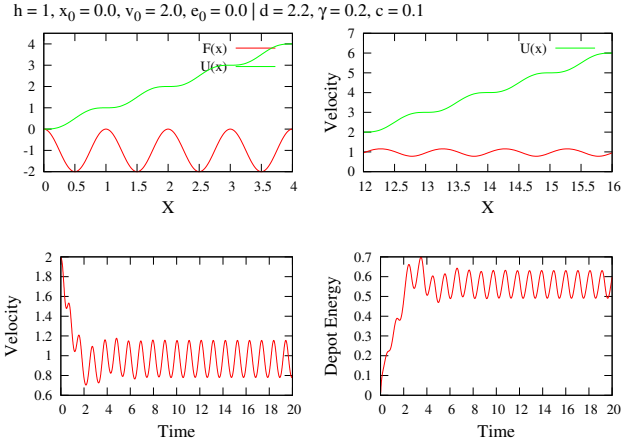}
  \caption{Active motion on staircase ratchets for a rather large value of
  the strength of driving $d = 2.2$.}
  \label{bier1}
 \end{center}%
\end{figure}

\begin{figure}[htbp]
 \begin{center}
  \includegraphics[angle=-90, width=8.5cm]{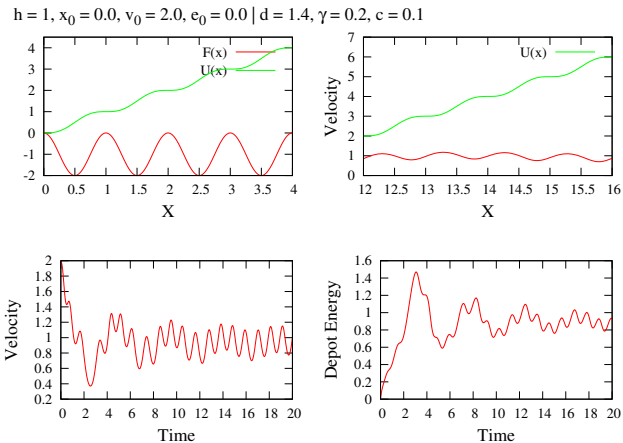}
  \caption{Active motion on staircase ratchets for a smaller value of the driving which
  is near to the threshold of uphill motion $d=1.4$.}
  \label{bier2}
 \end{center}%
\end{figure}

Among various ratcheting devices, a special class are so called staircase ratchets. This
type of ratchets may find applications to understand and design biological stepmotors
\cite{Bier03,Bier07}. Usually stepmotors have two phases:
(i) a power stroke where the legs move against a force and
(ii) a phase of free motion or diffusion. We propose to
model those by a staircase ratchet having a steep and a
flat region. \\
As a simple example we study the potential
obtained by a sinusoidal ratchet with a very strong tilt
(See the first plot in Figs. \ref{bier1} and \ref{bier2}):
 \be U(x) = h \left[ x
- \frac{1}{2 \pi} \sin(2 \pi x) \right].
 \ee
Here the parameter $h$ denotes the height of one step. As can be inferred from
Figs. \ref{bier1} and \ref{bier2}, the readjustment of the
exchange energy parameter $d$ can drive the system uphill
in the staircase ratchet. Fig. \ref{bier2} presents a longer
stabilization time because we are close to the critical $d_{cr}$
value below which the uphill motion is not observed. In
fact, for $d=1.3$ we observe only a sliding down motion of the
system. Moreover, in the chosen potential the initial velocity plays a key
role. Initial velocities lower than the limit mean value
(equal to $1$ in Figs \ref{bier1} and \ref{bier2}), give
rise to slopping down motion only.

\begin{figure}[htbp]
 \begin{center}
  \includegraphics[angle=-90, width=8.5cm]{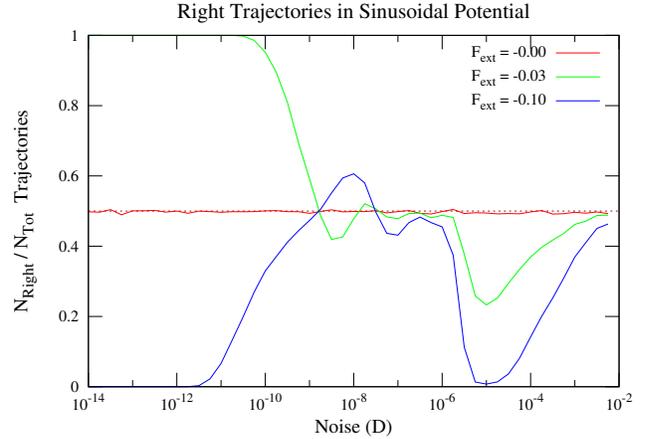}
\caption{Statistics of the right/left trajectories for the
 sinusoidal ratchet with and without the presence
different constant forces as a function of noise-intensity.
}
  \label{num-sin-traj}
 \end{center}%
\end{figure}

\begin{figure}[htbp]
 \begin{center}
  \includegraphics[angle=-90, width=8.5cm]{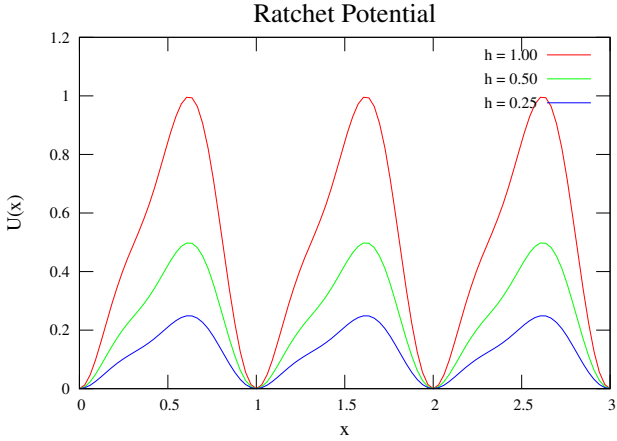}
  \caption{Ratchet potential obtained adding two sinusoidal functions for different height $h=1.0, 0.5, 0.25$ (Eq. \ref{pot}).}
  \label{pot_mat}
 \end{center}%
\end{figure}

As a last result for the sinusoidal-like ratchet, we present
numerically evaluated statistics of the right trajectories as a
function of the noise intensity for both: the tilted and
tilt-free ratchet. In Fig.\ref{num-sin-traj} we plot the
case of sinusoidal potential with and without additional
constant force in the equation motion. As expected, different
white noise intensities don't lead to any asymmetry in the
statistics of the flux directionality (right or left) in
the absence of the constant force ($F_0=0$). In contrast, the
additional constant force, strongly affects the motion.
 For very low values of noise, such as in the
deterministic regime, we observe that superposition of
both forces added gives rise to different flux
directionality for the set of parameters used.

When increasing the noise
intensity in the region $D \approx [10^{-9}, 10^{-6}]$, a
non-monotonic behavior of the fraction of right-oriented trajectories $N_{Right}/N_{Tot}$
as a function of the noise intensity
is observed. By further increasing  the noise $D
\approx [10^{-6}, 10^{-3}]$, a region of strong preference towards
left-oriented  trajectories is detectable, giving the expected behavior
of the flux in agreement with the applied constant force.
Still higher values of noise intensity blurred the picture and the rectifying
propertie sof the system  are no longer detectable ($N_{Right}/N_{Tot}\approx 0.5$).

This \textit{stochastically driven directionality} of the sinusoidal
ratchet system in the presence of an additional
tilt appears to be a generic effect and will be further discussed (for asymmetric ratchets) in the forthcoming
section,
 where similar statistics of $N_{Right}/N_{Tot}$ events has been
evaluated.

\section{Asymmetric Smooth Ratchet Potential}

\subsection{Analytical results from perturbation theories}

As in the case of the sinusoidal potential, also for the
asymmetric ratchet case we observe three different regimes depending
on the depot-energy influx to the
mechanical motion:\\
(i) the state of complete rest (point attractor of the dynamics),\\
(ii) sustained oscillations in one well (bounded attractor),\\
(iii) flux regime (open attractors).\\
The bifurcation plot of these regimes is isomorphic to that one
of the sinusoidal potential and it is shown in
Fig.\ref{bifurcation}. We study here only the flux regime
which is the most interesting regime for rectifying  ratchet systems. The
analytical expressions for the trapped cases (
corresponding to the harmonic approximation of the potential),
are identical to those reported in the previous chapter.

\begin{figure}[htbp]
 \begin{center}
  \includegraphics[angle=-90,width=8.5cm]{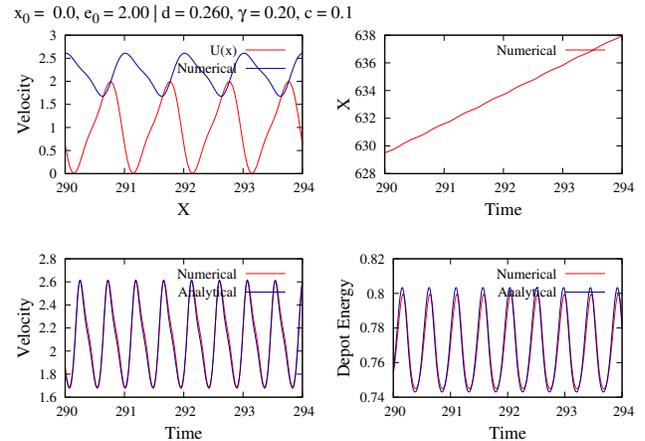}
\caption{Asymptotic behavior of the system in the flux case
with the usage of the Mateos asymmetric ratchet potential
plotted in Fig.\ref{pot_mat}.}
  \label{fflux}
 \end{center}%
\end{figure}

Examples of trajectories in the flux regime are given in
Fig.\ref{fflux},\ref{examples}. By inspection of these
plots we conclude that the change in the right and left asymptotical
velocity is not monotonically dependent on the energy
transfer parameter $d$.

The analytical evaluation  of the velocity and the energy
in the ratchet case, follows the same scheme used for the
sinusoidal potential. Using the expression of the velocity
derived in the perturbation treatment (Eq. (\ref{approx})),
with the ratchet of the Eq.(\ref{pot}), we can proceed with
the evaluation of the analytical asymptotic behavior of the
depot energy using the expression of the velocity written
in the following way:
 \be
   \dot{x}_{\infty}(t) = b + a \sin(\omega t)+ f \sin(2\omega t),
 \label{feq3}
 \ee
the formal analytical expression for the $e(t)$ is then:
 \be
   e_{\infty}(t) = [qY(t) +  e_0 e^{-db(f+2a)/\omega})] e^{-\zeta(t)}
 \ee
where
 \be
   Y(t)=\int_0^t e^{\zeta(t')} dt'
 \label{feq5}
 \ee
and
 \be
   \zeta(t)=(c + d
   b^2+\frac{da^2}2+\frac{df^2}2)t-\frac{\theta(t)}{\omega},
 \ee
with
 \bea
   \theta(t)&=& 2dba \cos(\omega t)-daf\sin(\omega t) +  \nonumber \\
            &+& dfb\cos(2\omega t)+ \frac{1}4 da^2 \sin(2 \omega t) + \nonumber \\
            &+& \frac{1}3 dfa \sin(3 \omega t)+\frac{1}8 df^2 \sin(4 \omega t).
 \eea
Because the term $(c + d b^2+da^2/2+df^2/2)$ is again
greater than zero, the equation has a non vanishing and non
diverging asymptotic behavior. The asymptotic expression we
can derive from the equation is then:
 \begin{equation}
  e_{\infty}(t)=\frac{q}{c + d b^2+da^2/2+df^2/2}
e^{\theta(t)/\omega}.
 \label{asy-ene-1}
 \end{equation}
The expression for $e_{\infty}(t)$ contains in the
exponential six oscillating terms with four different
frequencies ($\omega$, $2\omega$,$3\omega$, and $4\omega$).
The relative weight of the higher frequencies are low
because of the coefficient of these terms and also because
the value of the coefficient $f$ which is equal, in our
potential, to one forth the coefficient $b$. The relevant
oscillating term is then that one with frequency $\omega$,
the others representing less significant harmonics.\\
Performing the average of the asymptotic energy $\langle
e_{\infty}(t) \rangle_{\tau}$ as made with the sinusoidal
potential, and equalizing it to the value $\gamma / d$, we
can simplify the expression for the energy $e_{\infty}(t)$
as:
 \be
  e_{\infty}(t)=\frac{\gamma}{d} e^{\theta(t)/\omega}.
 \label{asy-ene-2}
 \ee
that is the expression used in our predictions.
Fig.\ref{fflux} shows the behavior of this expression
compared with that one obtained by numerical evaluation.
Even in this asymmetric case, the agreement is extremely
good. The parameters $a$, $b$, $f$ come from the expression
of the velocity, and $\omega$ has been extracted from the
numerical evaluation of the limiting velocity $v_{\infty}(t)$.
Otherwise, $\omega$ can be also predicted by means of the mean flux
velocity $\langle v_{\infty}(t)\rangle_{\tau}=b$ using the
relation:
 \begin{equation}
   b T = \frac{b}{\nu} =\frac{2 \pi b}{\omega} =  L =1 \Rightarrow \omega = 2 \pi
   b,
 \end{equation}
where $T$ is the period of the oscillations, $\nu$ the
corresponding frequency and $L$ is the length of the
ratchet periodicity.

\subsection{Results of simulations}
Numerical simulations performed by Tilch et al. (1999) have documented that, at least for
 ratchet models with piecewise linear potentials, the onset of a directed net current appears in two different directions.
Similar  behavior is registered in this study: at sufficiently large values of driving, close to the stationary states $v_0^+$ and $v_0^-$, the dynamical system
possesses open attractors corresponding to the left or right
current states.
\begin{figure}[htbp]
 \begin{center}
\includegraphics[angle=-90,width=8.5cm]{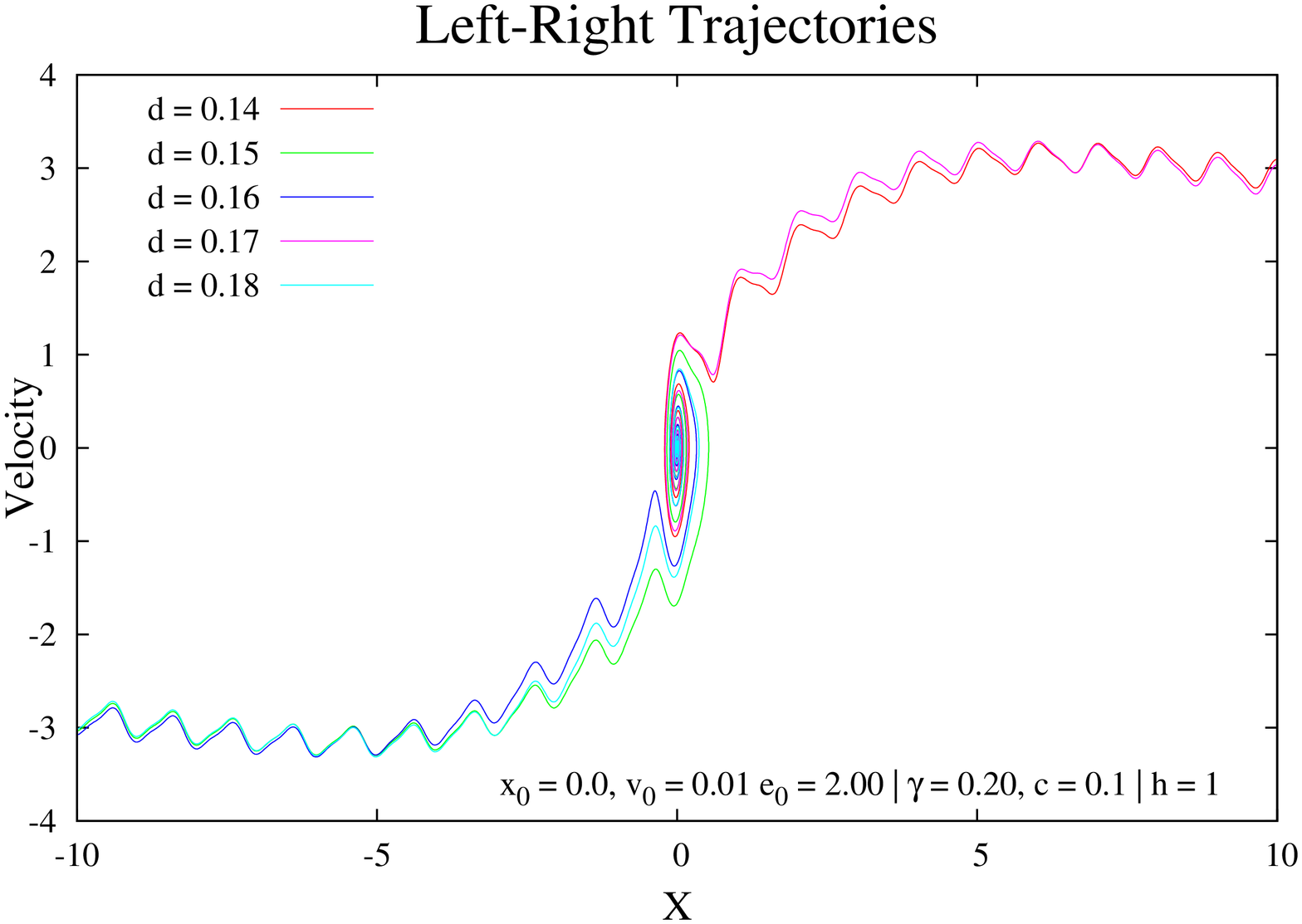}
\includegraphics[angle=-90,width=8.5cm]{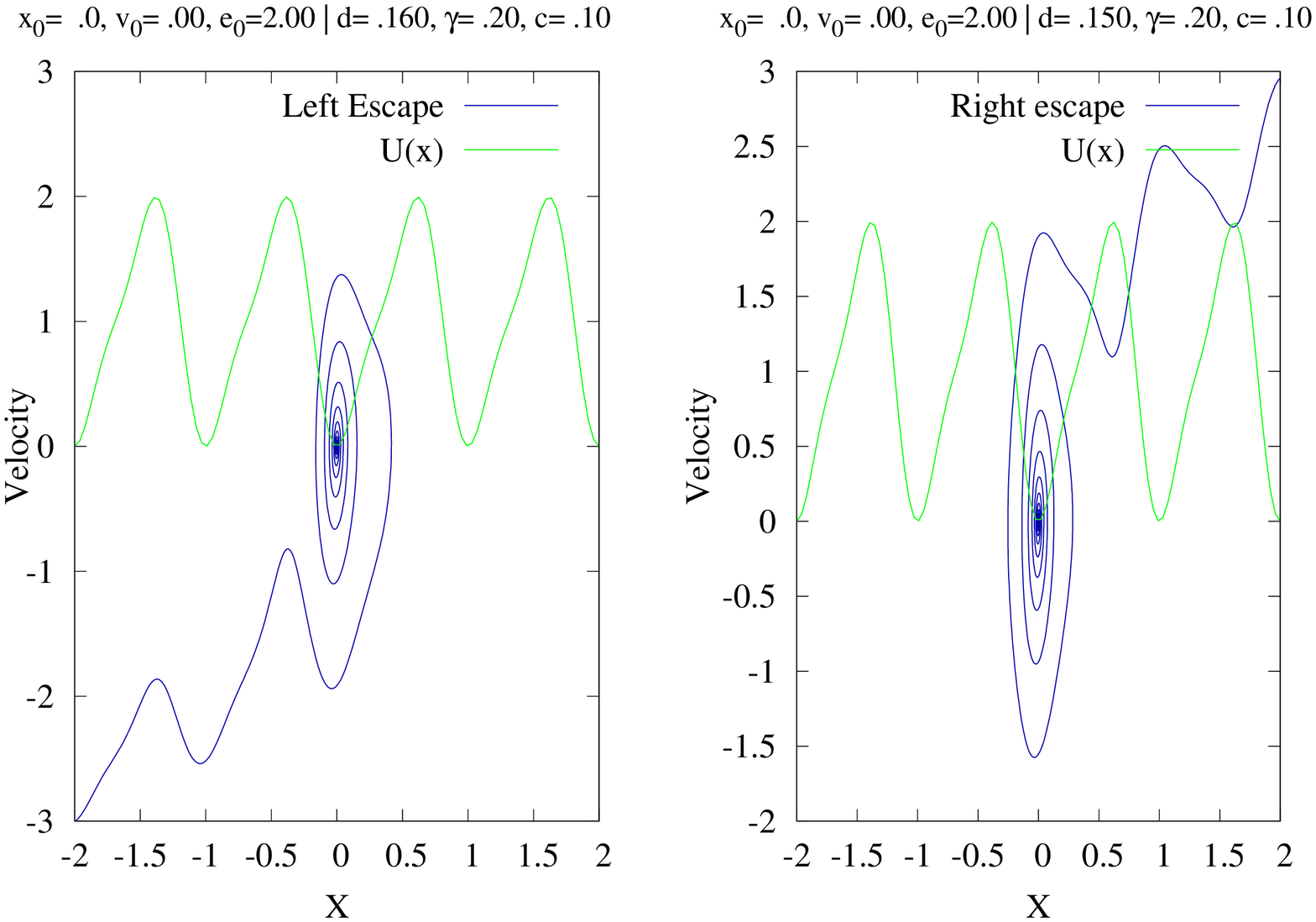}
\caption{Examples of escape events from the well of
the ratchet-potential. Relatively small variations of the energy
transfer parameter $d$ can induce different
directionality of flux.}
  \label{examples}
 \end{center}%
\end{figure}
As we can see in Fig. \ref{examples}, the ratchet driven by
active friction with a strong depot-particle coupling ($d>
d_E$) possesses 2 momentum-dependent attractors.
Slight variations of the intensity  $d$ result in changes of the flux directionality, especially in the domain of low noise-strengths (cf. Fig. 15) where the dynamics follows the trend observable in a deterministic system. The overall pattern, however, does not reveal any systematic tendency.
\begin{figure}[htbp]
 \begin{center}
  \includegraphics[angle=-90,width=8.5cm]{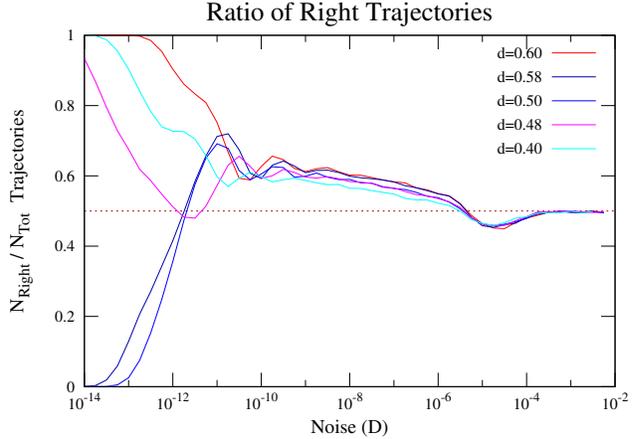}
\caption{Fraction of right-directed trajectories $N_{Right}/N_{Tot}$ out of the total number of analyzed cases $N_{Tot}=5\times 10^4$.
 Directionality of the
asymptotic velocity is displayed as a function of the noise intensity.
}
  \label{Num-asym-traj}
 \end{center}%
\end{figure}
In case of stochastic dynamics, two asymptotical
velocities  can be compared with
 two modal values (most probable ones) of the stationary probability density $P_0(v)$.
Here, instead of examining positions of maxima of $P_0(v)$, we have analyzed portion of trajectories whose long time evolution results in a preferred direction of motion (Fig. (\ref{Num-asym-traj})).
Closer inspection of Fig. (\ref{Num-asym-traj}) allows to detect noise-induced changes in the directionality of current.
For very
low noise intensity, a deterministic scenario prevails and the direction of current is decided by the deterministic dynamics
(cf. Fig.(\ref{examples})). In contrast, at increasing values of the noise intensity
($D \simeq
[10^{-12},5 \cdot 10^{-4}]$) the system exhibits a tendency to move towards right.
At still higher values of $D
 \simeq [5 \cdot
10^{-4},10^{-2}]$, the preferred asymptotic direction
changes to the left. Eventually, for values of $D \simeq 10^{-3}$ the motion of the system becomes fully delocalized resulting in equal ratios of
 trajectories going to the left and right.

\subsection{Tilted ratchets}

First we discuss the existence of an unidirectional ratchet.
Studying a Mateos ratchet it is allowed a choice of parameters
such that the average of the smaller slope (increasing left
to right) can still be overcome by the driving mechanism.
However the large slope (from right to left) is too large
to be overcome. In other words, there exists an uphill
solution for the smaller slope and no uphill solution for
the larger slope. This prevents any possibility to go left
in our case. The ratchet-system is acting then as a unipolar, rectifying device.

Under tilted ratchets we understand ratchets with a
constant average slope. In other words we have a global
incline of the ratchet which is due to some constant
average force. This may model a constant external load
against which the ratchet has to do work. We mention that
several authors considered also the case of
oscillating tilts.
We are interested in doing work against a load, therefore
consider here only constant external slopes.
\begin{figure}[htbp]
 \begin{center}
  \includegraphics[angle=-90,width=8.5cm]{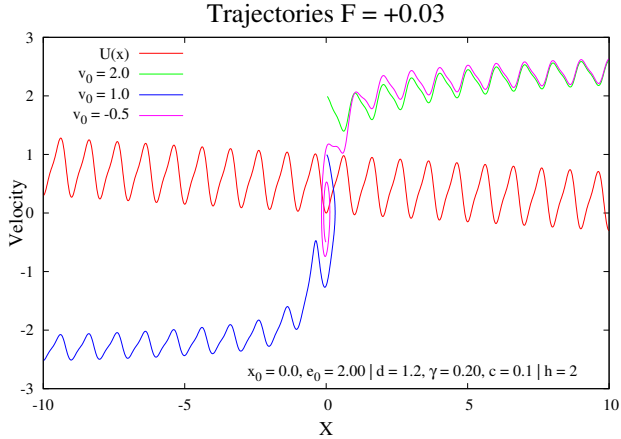}
\caption{Examples of uphill flux on tilted ratchets. The
load force is directed left to right. }
  \label{tiltrat9}
 \end{center}%
\end{figure}

In Fig. \ref{tiltrat9} we give an example of an uphill
motion against a force (here F = 0.03). The load force is
in the example directed left to right. We see that in
dependence on the initial conditions both directions may be
realized.

\section{Applications}

The investigated systems may be of interest for the modelling of
molecular motors which are able to convert chemical energy into
mechanical energy (work). We see possible applications to proton
pumps, the rotating
motors connecting with the work of ATP-ase \cite{Jul95,Jul97}
and also to step motors as proposed by Bier \cite{Bier03,Bier07}.

Molecular motors are nanotechnological objects being the
result of biological evolution. All molecular motors use
the energy quantum connected with the synthesis or
hydrolysis of the nucleotides ATP/ADP or the difference of
the electrochemical potentials on the cytoplasm membrane.
Standard models of molecular motors are based on the Smoluchowski
equations for discrete systems having several states which
correspond to attachment or detachment \cite{Jul95,Jul97}.
Many models have been developed
which follow similar lines. We follow in this work another
route which is based on Hamiltonian ratchets. We studied
Hamiltonian ratchets which are connected to an energy
reservoir and gave special attention to possible applications
to molecular motors.
We investigated in detail the motion of a particle
against a gradient of the potential i.e. uphill motion under
conditions where the external force is pointing downhill. The
general schema is the following: chemical energy is absorbed in
the form of ATP and introduced into our "motor" increasing the
reservoir of $e(t)$ by a certain amount. This is modelled here
by a continuous inflow $q$. In some other work we develop
a more refined model based on the assumption of discrete energy quanta,
representing the absorption of one molecule ATP \cite{EbGuFiAPPB}.
The absorbed energy flows to the "motor" and is
transformed into mechanical or electrical potential energy. This could
e.g. model the increase of the energy of protons by transport through
the membrane.

\begin{figure}[htbp]
 \begin{center}
  \includegraphics[width=8cm]{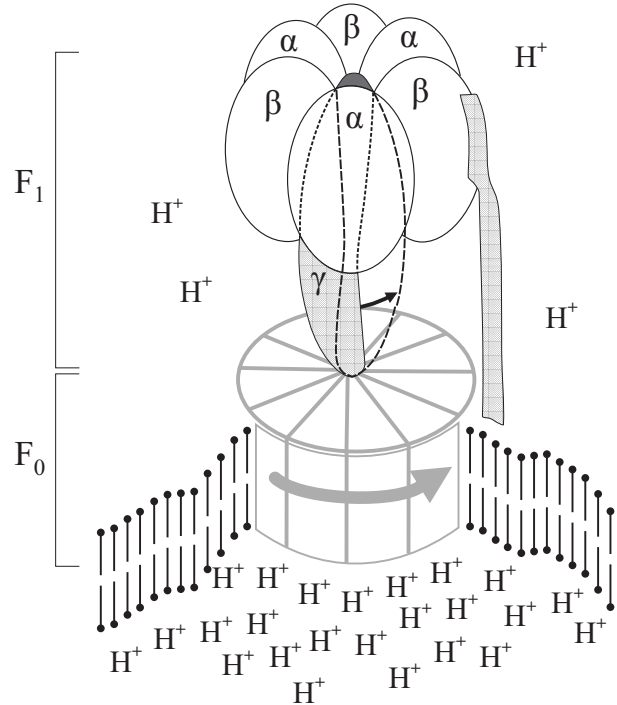}
  \caption{Scheme of the molecular motor $F_1 F_O$-ATPase-synthase. The complex $F_o$ which is
  imbedded into the membrane is responsible for the transport of protons through the
membrane. The complex $F_1$ sitting on the membrane is
responsible for the catalytic functions during the
synthesis / hydrolysis of ATP. The figure has been adapted
from an online publication of the Royal Swedish Academy of
Sciences
(\textit{http://nobelprize.org/nobel\_prizes/chemistry/laureates/1997/ press.html}).}
  \label{F1F0}
 \end{center}
\end{figure}

In the case of the ATP-ase motor,
in one of the direction of its rotational kinematics, the enzyme
ATP-ase hydrolyzes ATP into ADP and anorganic phosphate-Pi,
releasing energy and moving protons. The work of $F_OF_1$-ATP-ase is
connected with the rotation of a "rotor". The idea about  a
"rotor" corresponds to our knowledge about the structure
(see Fig. \ref{F1F0}).
In our model of active Brownian particles the variable $x$ has to be interpreted then
as the rotation angle of the rotor. \\
Another possible application may be to model step motors, as studied e.g. by
Bier \cite{Bier03,Bier05,Bier07}. Here the motion of molecular
legs is composed by two phases. The first phase is the
power stroke where the system works against a force, and
the second phase is the free or diffusive motion. One possibility
to model such two-phase motions is to use staircase-like ratchets
defined by Eq.(52). Another possibility is to develop ratchet
models based on two active particles coupled by a spring.
Step motor models of this type, based on overdamped ratchet motion,
were first studied by Derenyi and Vicsek \cite{Vicsek}.

\section{Summary} In this work we analyzed a mechanical
system with inertia subjected to a dissipative forcing
having two terms: one passive damping and an 'active'
contribution providing energy to the system
by means of an energy supply
described in an additional coupled equation. The system has
been studied under the influence of
smooth ratchet potentials: symmetrical (sinusoidal
potential) and asymmetrical
(Mateos-type potential and tilted ones). The
analytical and numerical evaluation of the equations has
been found for both the asymptotical velocity of the
mechanical system and the depot energy. The system presents
bifurcations of the asymptotic velocity as a function of the
energy transfer parameter $d$. The three regimes found
correspond to: 1) relaxation in a potential minimum (vanishing motion); 2) oscillating motion in a well (limit cycle), and 3)
flux motion with two values of the asymptotic velocity. The
motion in the flux regime is then possible in two directions, even in the
presence of a tilt in the potential. The numerical simulations of the
system under the action of white Gaussian
fluctuations show the effect of noise-controlled directionality of the motion.

Possible applications of the
ABM system in modelling molecular motors connected to the $ATP$
synthesis/hydrolysis have been briefly discussed.

\vspace{1cm}
This work has been supported by the Marie Curie TOK grants
under the COCOS project (6th EU Framework Programme,
contract No: MTKD-CT-2004-517186) as well as the ESF programme STOCHDYN. The authors thank Martin Bier for discussions and advice.

\end{document}